\documentclass[a4paper]{jpconf}
\usepackage{graphicx}
\usepackage{amsfonts}
\usepackage{amssymb}
\usepackage{multirow}

\def\mn{{\mu\nu}}

\def\b{\beta}

\def\f{\phi}

\def\r{\rho}
\def\G{\Gamma}
\def\L{\Lambda}
\def\A{{\cal A}}
\def\B{{\cal B}}
\def\O{{\cal O}}
\def\S{{\cal S}}
\def\prd{Phys.\ Rev.\ D}
\def\plb{Phys.\ Lett.\ B}

\begin{document}
\title{Inflation in asymptotically safe $f(R)$ theory}

\author{Adriano Contillo}

\address{SISSA, Via Bonomea 265, Trieste, Italy\\
INFN, Sezione di Trieste, Italy}

\ead{contillo@sissa.it}

\begin{abstract}
We discuss the existence of inflationary solutions in a class of renormalization group improved polynomial $f(R)$ theories, which have been studied recently in the context of the asymptotic safety scenario for quantum gravity. These theories seem to possess a nontrivial ultraviolet fixed point, where the dimensionful couplings scale according to their canonical dimensionality. Assuming that the cutoff is proportional to the Hubble parameter, we obtain modified Friedmann equations which admit both power law and exponential solutions.
We establish that for sufficiently high order polynomial the solutions are reliable, in the sense that considering still higher order polynomials is very unlikely to change the solution.
\end{abstract}

\section{Introduction}

The idea that a perturbatively nonrenormalizable 
theory could be consistently defined in the UV limit at a nontrivial fixed point (FP),
often called ``asymptotic safety'', is theoretically very attractive, 
especially when applied to quantum gravity \cite{weinberg}.
Much progress in this direction has come from the direct 
application of Renormalization Group (RG) techniques to gravity. 
A particularly useful tool has been the non-perturbative 
Functional Renormalization Group Equation (FRGE) \cite{wetterich},
defining an RG flow on a theory space which consists of all diffeomorphism invariant
functionals of the metric $g_{\mu\nu}$. 
It defines a one parameter family of effective field theories with actions 
$\Gamma_k(g_{\mu\nu})$ depending on a coarse graining scale (or ``cutoff'') $k$,
and interpolating between a ``bare'' action (for $k\to\infty$)
and the ordinary effective action (for $k\to 0$).

The application of such concepts to a cosmological framework is based to the following logic \cite{rw1}.
If we want to study the quantum evolution of the cosmic scale factor,
we should in principle use the full effective action $\Gamma(g_\mn)$, 
which, as we mentioned above,
coincides with $\Gamma_k(g_\mn)$ in the limit $k\to 0$.
However, our knowledge of this functional is rather poor.
One way of gaining some traction on this issue is to observe 
that the Hubble parameter appears as a mass in propagators.
Thus, the contributions of quantum fluctuations with wavelenghts 
greater than $H^{-1}$ are suppressed.
As a result, the functional $\Gamma_k$ at $k$ comparable
to $H$ should be a reasonable approximation for the same functional at $k=0$.
We do not know $\Gamma_k$ much better than the full effective action,
but we can easily calculate the dependence of some terms in $\Gamma_k$ on $k$.
By doing so we effectively take into account nonlocal terms
that would be very hard to calculate otherwise.

Previous investigations along these lines, in particular \cite{bore},
have been based on the Einstein-Hilbert (EH) truncation.
It is important to establish that the results obtained there 
persist when further operators are included in the truncation.
We know that at the FP the coefficients of these terms are not
very small, but their presence does not seem to affect the values
of the cosmological constant and Newton's constant too much.
In other words, the FP that is is seen in the EH truncation
seems to be robust.
The question then is to see if this stability of the FP against 
the inclusion of new terms 
is reflected in the stability of the corresponding solutions.
This question is important because
the values of the couplings at the FP are fixed and as a consequence
there are no free parameters to be varied.
We will see that the (power-law or exponential) inflationary solutions 
are indeed stable against the inclusion of new terms, but establishing
this fact requires including a rather large number of terms.

\section{Asymptotic Safety}

It is well known that general relativity can be treated as an effective field theory. This means that it is possible to compute quantum effects due to graviton loops, as long as the momenta of the particles in the loops are cut off at some scale. The results are independent of the structure of any “ultraviolet completion”, and therefore constitute genuine low energy predictions of any quantum theory of gravity.

Given for a field $\f$ an effective action $\G(\f)$, it can always be written as a sum
\begin{equation}\label{infaction}
 \G(\f)=\sum g_i\O_i(\f)
\end{equation}
that runs over the infinitely many operators consistent with the undelying symmetries of the theory. Given an energy scale $k$, one can define \emph{dimensionless} couplings $\tilde{g}_i=g_i k^{-d_i}$, where $[g_i]=[k]^{d_i}$. The resulting picture is an infinite dimensional space parametrized by the $\tilde{g}_i$'s, where a Renormalization Group flow can be seen as a vector field that shifts any action along a trajectory
\begin{equation}
 \G(\f)\rightarrow\G_k(\f)=\sum\tilde{g}_i(k)k^{d_i}\O_i(\f)\;.
\end{equation}

The particular realization of RG that was used in this work is the one $\grave{a}\,la$ Wilson, based on the idea that the effective action describing physical phenomena at a momentum scale $k$ can be thought of as the result of having integrated out all fluctuations of the field with momenta larger than $k$. This can be achieved adding to the bare action $\S$ a term
\begin{equation}
 \Delta\S_k=\int dx\,\f(x)R_k\,\f(x)
\end{equation}
where $R_k$ is called the cutoff kernel (see \cite{cpr1} for details). This term acts as a mass for the fluctuations with momenta smaller than $k$. Following the usual procedure of background field method, we define a $k$-dependent generating functional of connected Green functions by
\begin{equation}
 e^{-W_k(J)}=\int D\f e^{-(S(\f)+\Delta\S_k(\f))-\int dx J\f}
\end{equation}
and the effective action as its Legendre transform
\begin{equation}
 \G_k(\f)=W_k(J)-\int dx J\f-\Delta\S_k(\f)\;.
\end{equation}

One can show that it obeys the so called functional renormalization group equation (FRGE)
\begin{equation}\label{frge}
 k\frac{d\G_k}{dk}=\frac{1}{2}\textrm{Tr}\left(\G_k^{(2)}+R_k\right)^{-1}k\frac{dR_k}{dk}
\end{equation}
where $\G_k^{(2)}$ is the inverse propagator of the graviton. From the r.h.s. of (\ref{frge}) one can then extract the $\b$-functions of the couplings, as it is done in \cite{cpr1}, and study the RG flow of the theory towards the UV limit $k\rightarrow\infty$.

\pagebreak
A theory is then said to be asymptotically safe if
\begin{itemize}
 \item for any coupling $\tilde{g}_i$ there exists a finite UV limit $\tilde{g}_i^\ast$ that is called a fixed point
 \item the stability matrix $M_{ij}=\left.\frac{\partial\b_{\tilde{g}_i}}{\partial\tilde{g}_j}\right|_\ast$ evaluated in the fixed point has a finite number of negative eigenvalues $m_i$ (in the following, it will come useful to define critical exponents $\theta_i=-m_i$).
\end{itemize}

If both the requests are satisfied one finds that the \emph{critical surface}, the locus of the points that are attracted towards the FP, is finite dimensional. This means that only a finite number of experiments is needed to univocally identify a trajectory, and the theory is thereby predictive.

\begin{center}
 \begin{figure}[ht]\label{FPthetas}
  \includegraphics[width=6.5cm,height=10cm]{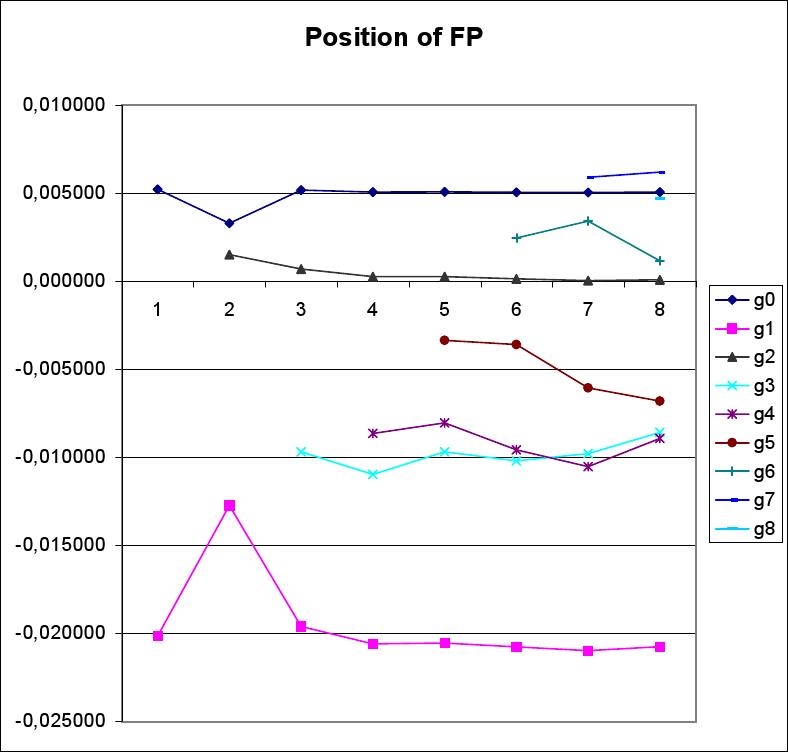}
  \includegraphics[width=6.5cm,height=10cm]{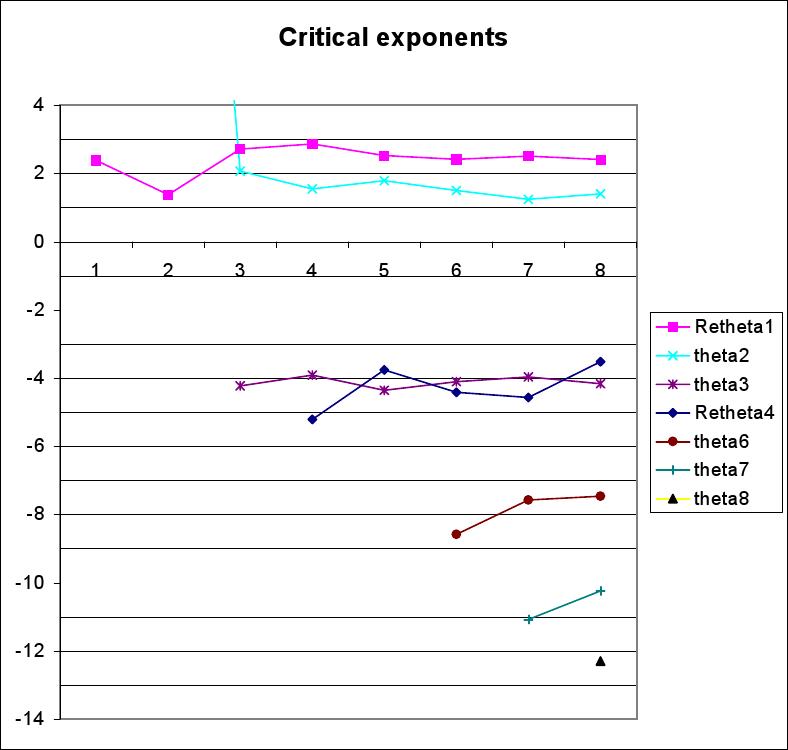}
  \caption{Position of the fixed point and values of the critical exponents as functions of the truncation order $n$.}
 \end{figure}
\end{center}

In \cite{cpr1} this procedure was applied to a (truncated) gravitational action consisting of a sum of powers of the curvature scalar $R$
\begin{equation}\label{gravaction}
 \G(g)=\int d^4x\sqrt{|g|}\,\sum_{i=0}^n g_iR^i
\end{equation}
with $n=2,\dots,8$, and it was further extended to $n=9,10$ in \cite{bcp}. For every $n$ a fixed point was found, with only three positive critical exponents. To understand if a finite truncation is a reliable approximation of the action (\ref{infaction}), one can study the stability of the position of the fixed point {$\tilde{g}_i^\ast$} and of the values of the critical exponents $\theta_i$: these values should stay almost constant with respect to the inclusion of new operators in the action. In figure (\ref{FPthetas}) it is clearly visible that this seems to be the case, so that our finite sum is a reliable truncation.

\section{Asymptotically safe cosmology}

It is now time to apply the above described formalism to a cosmological background. There are two main issues that are worth stressing, the first concerning the possibility to obtain any inflationary solution, and the second about the resulting stability of such solutions with respect to the inclusion of new terms in the action. We will address them after having defined a key ingedient of our analysis: the identidication of the RG scale $k$ with a characteristic scale of the described system.

As it was said before, in the Wilsonian approach the cutoff scale is the momentum scale at which the process is taking place. This means that, in a Friedmann-Robertson-Walker metric, it should be set to the order of magnitude of the Hubble rate $H(t)$. Another argument can be found in the comparison with the framework of quantum field theory in curved spaces, where the Hubble rate appears as a mass term in the propagator of the field. We then set $k=\xi H$, where $\xi=\O(1)$ is a free parameter that in principle can be tuned to offset the scheme dependence carried by the cutoff scale.

\subsection{Inflationary cosmology}

We have now all the ingredients to extract the desired inflationary solutions. From the action
\begin{equation}
 \G(g)=\frac{1}{16\pi G}\int d^4x\sqrt{|g|}\,(f(R)-\L)
\end{equation}
where $f(R)=\sum f_i R^i$, we derive the modified Einstein equations that, after having imposed the FRW symmetry, appear as (modified) Friedmann equations
\begin{eqnarray}
 \A(H)&=&8\pi G\r+\L\\
 \B(H)&=&8\pi G\r(1-3w)+4\L
\end{eqnarray}
where $w$ si the equation of state of the ordinary matter and $\A(H)$, $\B(H)$ are two functions of the Hubble rate whose shape depends on the chosen $f(R)$, and can be rearranged in the form
\begin{eqnarray}\label{fried2}
 \B(H)&=&(1-3w)\A(H)+3(1+w)\L\\
 \r&=&\frac{1}{8\pi G}\left(\A(H)-\L\right)\;.
\end{eqnarray}

Before solving the equations, we have to perform the above mentioned cutoff identification: the couplings are rewritten as
\begin{equation}
\Lambda=\tilde\Lambda k^2\ ;\quad
G=\tilde G/k^2\ ;\quad
f_i=\tilde f_i k^{2-2i}
\end{equation}
and, because we are interested in describing the dynamics of the very early universe, are assumed to lie in the UV fixed point. With these substitutions, we look for solutions of the (\ref{fried2}) in the form of a de Sitter ($H(t)=\bar H$ and $a(t)\propto e^{\bar Ht})$) and a power law ($H(t)=p/t$ and $a(t)\propto t^p$). It comes out that the system admits the former only for discrete values $\xi_{dS}$ of the free parameter, while we found continuous solutions of the latter type, with $p=p(\tilde\Lambda^\ast,\tilde G^\ast,\tilde f_i^\ast,w,\xi)$. As a final remark, the functions $p(\xi)$ show vertical asymptotes, located exactly at the values $\xi_{dS}$: de Sitter is then recovered as a limit of power law.

\begin{center}
 \begin{figure}[ht]\label{neq8}
  \includegraphics[width=8cm,height=6cm]{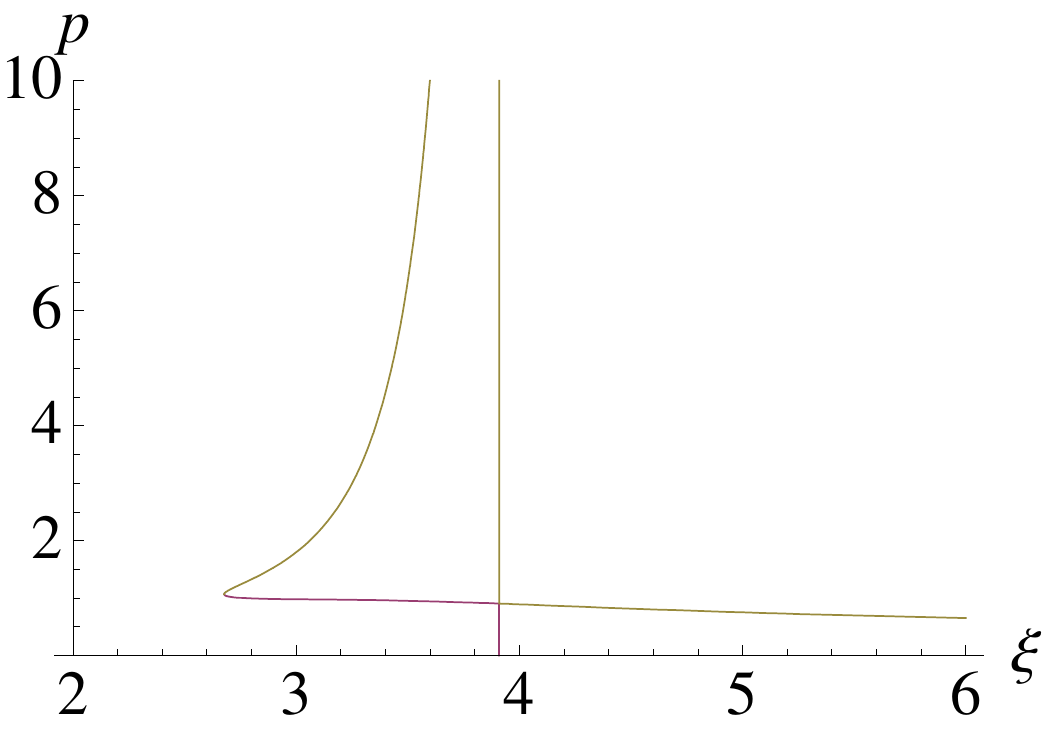}
  \caption{Power law exponents as functions of the parameter $\xi$. The asymptote is clearly visible for $\xi\simeq3.9$.}
 \end{figure}
\end{center}

In figure (\ref{neq8}) the power law exponents $p$ obtained for $n=8$ are plotted as functions of $\xi$. There is in particular one growing solution that is greater than 1 (\emph{i.e.} $\ddot{a}/a>0$) for most values of the free parameter.

\subsection{Reliability analysis}

As it was said before, the UV fixed point of the gravitational action (\ref{gravaction}) tends to stabilize as one widens the truncation. The issue is then whether the values of the solutions will show the same behaviour. We thus define a reliability criterion based on the following logic: taken the dimensionless curvature scalar $\tilde R=Rk^{-2}$, we write the dimensionless action with truncation $n$ evaluated at the  fixed point $\left.\tilde f(\tilde R)\right|_n$ and we compare it with the same object in $n-1$. A truncation is then said to be reliable if
\begin{equation}
 \left|\frac{\left.\tilde f(\tilde R)\right|_n-\left.\tilde f(\tilde R)\right|_{n-1}}{\left.\tilde f(\tilde R)\right|_{n-1}}\right|<5\%
\end{equation}
and in general this is only true for a limited range of values of $\tilde R$. One can schematically rewrite this as $\tilde R<c$, where

\begin{center}
\begin{tabular}{|c|r|r|r|r|r|r|r|r|r|r|}
 \hline
 $n$ & 1& 2& 3& 4& 5& 6& 7& 8& 9& 10\\
 \hline
 $c$ & 0.40& 0.24& 0.45& 1.07& 1.21& 0.90& 0.79& 0.92& 1.09& 1.09\\
 \hline
\end{tabular}
\end{center}

The table shows that the reliable range tends to increase, so that a larger truncation is more likely to correctly approximate the infinite truncation $n\to\infty$.

\begin{center}
 \begin{figure}[ht]\label{rel}
  \includegraphics[width=7cm,height=5.25cm]{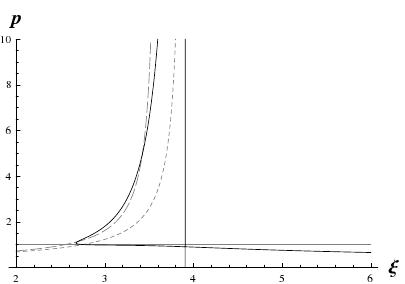}
  \includegraphics[width=7cm,height=5.25cm]{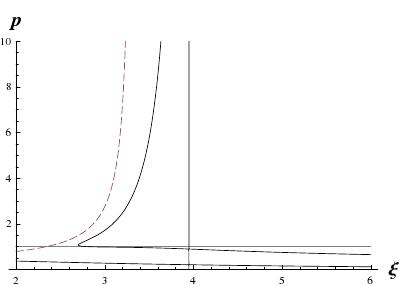}
  \caption{Power law exponents as functions of $\xi$ for $n=8$ (left) and $n=10$ (right). Areas to the right of the dashed lines are the reliability regions, while the dotted line indicates the $n=1$ solution described in \cite{bore}.}
 \end{figure}
\end{center}

For a power law solution, the constraint on the dimensionless curvature scalar can be translated into a constraint on the values of the exponents, namely
\begin{equation}
 c\xi^2\gtrsim6\left(2+\dot{H}/H^2\right)=6\left(2-1/p\right)
\end{equation}
that can be plotted together with the exponents themselves, as it is shown in figure (\ref{rel}) on the left. From the plot it comes clear that only a small portion of the growing solution is to be considered reliable, so we decided to extend the truncation to $n=10$. The result is plotted in figure (\ref{rel}) on the right, that shows a great improvement in the reliability of the solutions.

\section{Conclusions}

The results obtained here can be summarized saying that, if we identify the cutoff with a multiple of the Hubble parameter, inflationary power law solutions (i.e. with exponent $p>1$) exist for some range of values of $\xi$. The dependence on such parameter is strong, with the exponents diverging at some values $\xi_{dS}$. Exactly at those points, the theory admits also de Sitter solutions. Provided that $\xi$ lies in the above range, that the starting point is close enough to the FP and that $p>1$, it should always be possible to have a sufficient number of $e$-foldings.

Moreover, it has been shown that the increasing stability of the fixed point with respect to the widening of the truncation seems to be reflected in an increasing reliability of the cosmological solutions obtained using the truncated action. The overall conclusion that one can draw is that we found a strong hint of the convergence of the series in our fixed point action, so that any result coming from low order truncations should not be too misleading.

\end{document}